\documentstyle[psfig]{article}
\begin{document}
\def\astrobj#1{#1}

\title{Rapid Variations of Narrow Absorption Line Components 
in the Spectrum of \astrobj{P Cyg}ni}

\author{Cornelis de Jager{\footnote{e-mail: C.dejager@sron.nl}} \\ 
SRON Laboratory for Space Research \\
Sorbonnelaan 2, 3584 CA Utrecht, The Netherlands
\and
Garik Israelian \\ Instituto de Astrofisica de Canarias \\
Via Lactea s/n, E-38200, La Laguna, Tenerife, Spain}

\maketitle

\begin{abstract}

We report the detection of rapid wavelength and intensity variations of 
narrow 
components in absorption line profiles of the hot galactic 
supergiant \astrobj{P Cyg}ni. During most of the time, in one week of observations, there 
were two such components present. Although the period of observation is too
short for firm conclusions, the velocity curves do not contradict the 
photometic period of 17.3 d. These curves may be interpreted as a double 
wave, 'swinging in counterphase'. This observation  
would suggest the presence of one long wave over the stellar 
surface. The waves are possibly gravity waves with wavelengths 
of the order of half the stellar radius. 

\vspace{0.3cm}

\end{abstract}

\section{Introduction; Long-Period DACs}

\astrobj{P Cyg} is a blue hypergiant (de Jager 1998) or a Luminous
Blue Variable (LBV, Humphreys \& Davidson 1994) whose observed 
characteristics make 
the star among the most peculiar ones in the Galaxy. The properties
of \astrobj{P Cyg} and its wind have been reviewed by Israelian and de Groot 
(1999). 

Narrow absorption line components called DACs were discovered in Balmer line
profiles of \astrobj{P Cyg} by de Groot (1969). He also found them in lines of 
He, Fe and many other elements. Identifications of other lines were 
published by others. Cassatella et al. (1979) and Luud and Sapar (1980)
detected DACs in UV lines of various ions, such as Cr\,{\sc ii}, 
Ni\,{\sc ii}, and others. Lamers et al. (1985) found them in 
Fe\,{\sc ii} lines.

In his Balmer line observations de Groot found velocities of
--215, --160 and --95 ${\rm km}{\,}{\rm s}^{-1}$. The first component 
showed periodic $v_R$-variations of $P =$ 114 days, while the others 
did not present temporal variations. The observations by Luud et al. (1975)
however, give contrasting results. They found that the first component 
was not variable, while the second was, with a period of 57 d. Since,
many characteristic times of variability (we prefer this term above 
the too suggestive 'period') have been mentioned: 200 d (Markova, 1986),
60 - 75 d (van Gent and Lamers, 1986), 4 to 5 months (Markova, 1993) 
and 200 d (Israelian et al., 1996). The apparent diversity in the 
communicated 'periods' may partly be related to insufficient lengths
of periods of observations; moreover, it may be a real phenomenon.  

As to the interpretation of these motions, current thinking converges on a 
model according to which shells are periodically ejected from the star, 
while increasing their velocity on the way outward through the stellar 
wind. This hypothesis of expanding shells was proposed first by Kolka 
(1983), Markova and Kolka (1984) and Markova (1986). They were 
supported by van Gent and Lamers (1986), and are also shown in a 
diagram published by Israelian et al. (1996), where three or four 
shells appear, of which the velocity increases steadily from about 80 
to 190 ${\rm km}{\,}{{\rm s}^{-1}}$ over a period of about 1000 days.

Most of the above mentioned variations have characteristic times of the 
order of 2 to 7 months. There is another component, though, the period of 
which is not related to such presumed outgoing shells. De Groot et al. 
(2001) published the results of a fairly complete survey of \astrobj{P Cyg}
photometry in which they demonstrate that at each time there 
are two periods in the light variation. The shorter one,
of 17.3 d, occurred persisitently, at least during the last 20 years
albeit with variable amplitude. The longer period ranges between 
about 60 and 130 d, with a clear maximum near 100 d. It does not seem 
impossible that this latter variability is associated with the 
ougoing shells described above. But what to say about the shorter period ? 
The question arises if there is any wavelike phenomenon in the wind that 
could 
be associated with the shorter (17.3 d) light variation. Its study is 
the main content of this paper.

\section{The detection of short-period absorption components}

Observations of \astrobj{P Cyg} were carried out during the week 1999,  
May 28--June 4, using the SOFIN Echelle Spectrograph of the 2.6-m 
Nordic Optical 
Telecope (NOT) at the ORM (La Palma). A resolving power of 
$R=\lambda /\Delta\lambda\sim$ 80\,000 has been achieved with 
the second camera. The CCD used in this run was a 
$1152 \times 770$ pixel$^{2}$ EEV. We obtained echelle 
spectra with a slit width of 0.8 arcsec. They cover the wavelength 
range between 3600 and 4200 \AA\ in 18 orders. All spectra were 
flat-fielded, background and scattered light subtracted. 
The wavelength calibration was performed with a Th--Ar lamp.
The ultimate signal-to-noise of our spectra at 4000 \AA\ is about 200.

\begin{figure}
\vbox{\psfig{file=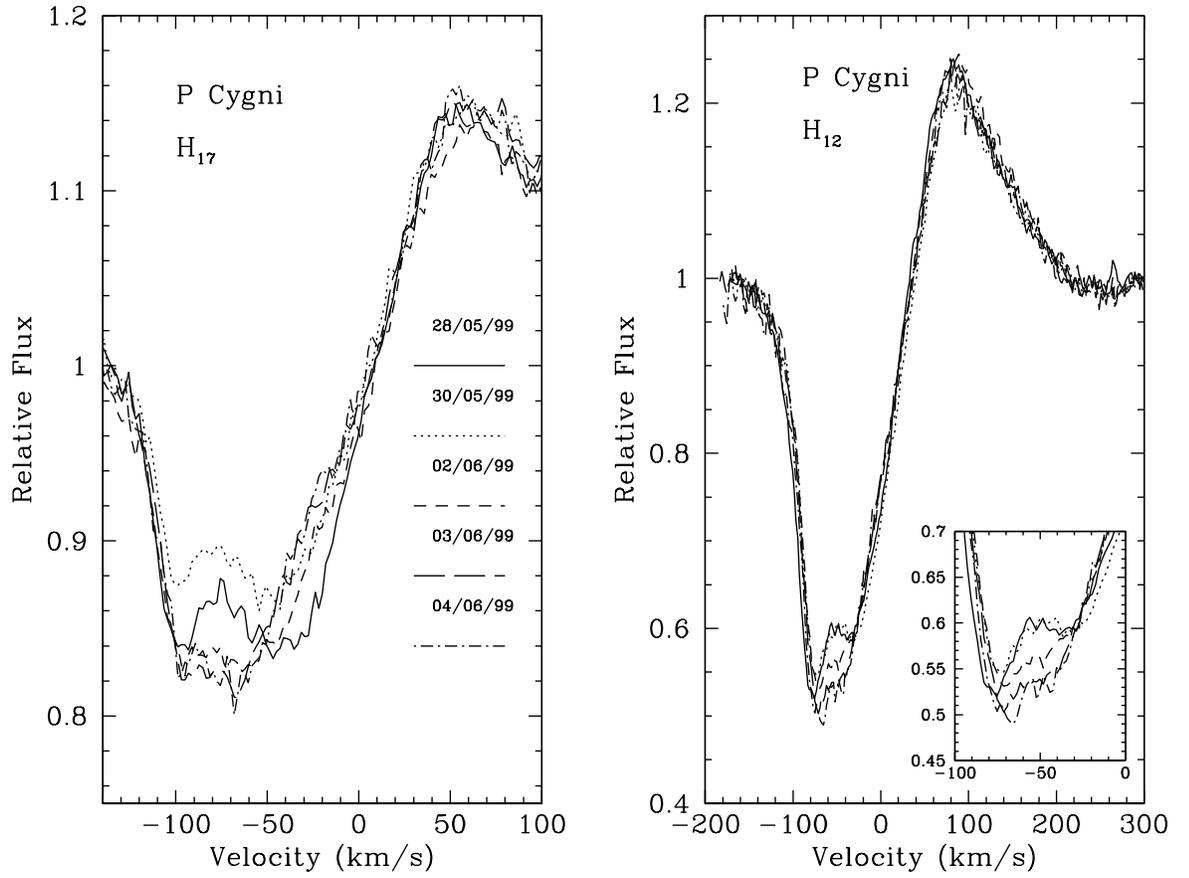,width=16.5cm,height=12.5cm,angle=270}}
\caption{High-resolution profiles of the Balmer lines ${\rm H}_{17}$ 
and ${\rm H}_{12}$ obtained with the NOT/SOFIN telescope combination
at the ORM (La Palma). During seven days of observations the profiles 
changed considerably.}
\end{figure}

Apparently, one week is too short for observing recurrent variability 
or for checking a periodicity of 17.3 d. In that 
respect our results have a preliminary character. The positive aspect
is, though, that our observations (Fig. 1) do show the presence of rapidly 
changing absorption components in profiles of the higher members of the 
Balmer series. 

The lines showed split absorption cores. These are best visible in the
profiles of 28/05/99 (solid lines in Fig. 1), which have two distinct 
cores. Closer examination shows that these cores existed also in later 
spectra. The cores approached each other during the next few days and 
nearly merged around 02/06/99. Thereafter they reappeared. We 
measured the 
wavelength variation of these absorption cores by two methods: 
directly by determining the central wavelengths of the absorption 
component, and also 
by fitting the profiles by two Gaussians. This was done in the two
lines, so we got 4 data points for each componentat each day. The 
measurements agree reasonably well; the mean scatter of the individual 
velocity measurements is 3 ${\rm km{\,}{s^{-1}}}$. Table 1 gives the
average velocities measured in the two line components.

\begin{table}
	\caption{Wavelengths of absorption components measured in the 
	profiles of 
${{\rm H}_{12}}$ and ${{\rm H}_{17}}$. The wavelengths $v$ are expressed 
in 
${\rm km{\,}{s^{-1}}}$. For details see the text.}
 
\vspace{0.2cm}

\begin{tabular}{|l|l|l|} \hline

date & $v$ & $v$ \\ \hline

28-05-99  & 87  & 49  \\
30-05-99  & 87  & 52  \\
02-06-99  & 75  & 72  \\ 
03-06-99  & 84  & 69  \\
04-06-99  & 80  & 66  \\ \hline

\end{tabular}
\end{table}

\begin{figure}
\vbox{\psfig{file=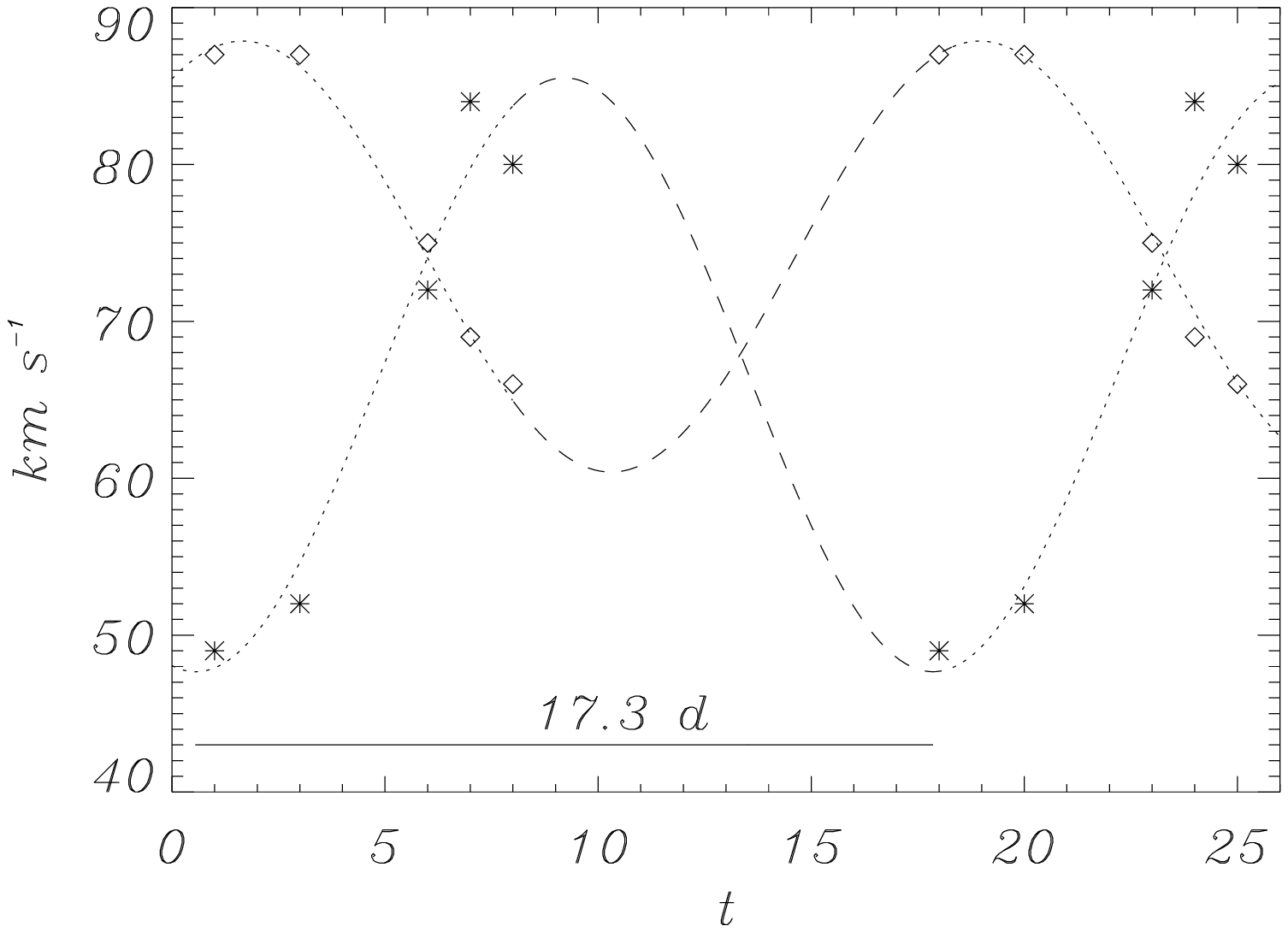,width=16.5cm,height=12.5cm,angle=0}}
\caption{Measured wavelengths for the two line components in the lines 
${\rm H}_{12}$ and ${\rm H}_{17}$. The data have been plotted for the
dates of observation and again for a period of time, 17.3 d later. 
Assuming
that the high-velocity component becomes the low-velocity one and 
vice-versa after a 'switch-over', the data may be interpreted as a periodic 
variation of two waves in counterphase; {\em cf.} the dotted lines.}
\end{figure}

We wanted to check if these velocities are related in some way to the 
photometric period of 17.3 d, reported in Section 1. To that end we 
replotted the data in the same diagram for a period, 17.3 days later 
(Fig. 2). The diagram  
shows that it is {\em not impossible} that the observations have a 
repetition period of 17.3 d, and it appears moreover that there are 
possibly 
two waves that are 'swinging in counterphase', {\em cf.} the dotted 
lines in Fig. 2. If this interpretation would be verified by 
measurements taken over a period of a month or longer, then they 
could be interpreted in terms of a long wave in the wind over the 
stellar surface having a wavelength of the order of the star's 
diameter. This hypothesis will be examined in the next section.   

An interesting aspect of the observations is the high acceleration
that is inferred from the wavelength variation of the absorption 
components. It is
about 10 ${\rm km}{\:}{{\rm s}^{-1}}{\:}{{\rm d}^{-1}}$. This 
value stands in contrast to the acceleration of the 
long-period DACs, which is about 100 times smaller, of the order of 
0.1 ${\rm km}{\,}{{\rm s}^{-1}}{\,}{{\rm d}^{-1}}$.

\section{A Possible Interpretation}

In this section we examine the following question: if a 
periodic motion with $P= 17.3$ d occured in the wind of \astrobj{P Cyg}, what 
would be the physical character and the wavelength of that motion?  

We base our answer on the dispersion equation for perturbations in
line driven winds as developed by Abbott (1980, in his eq. (45)). 
That equation reads

\begin{equation}
 \omega^{3} + \omega ^{2} k_2 F_2 - \omega ( a^2 {k_x}^2 + a^2 {k_z}^2) -a^2 {k_x}^2 k_z F_2 = 0 
\end{equation} 

with 

\begin{equation}
F_2 = \delta g_{\rm rad}/ (\delta (\delta w / \delta r))
\end{equation}

where $w$ is the wind velocity, $a$ that of sound, $k = 2 \pi / L$; $L$ is the 
wavelength, $\omega = 2 \pi/ P$ , where $P$ is the period of the 
disturbance and
 $g_{\rm rad}$ is the radiative acceleration. For purely horizontal or 
vertical waves $k_z = 0$ or $k_x = 0$ respectively.

For the calculation of the diagram we used the model of the wind 
of \astrobj{P Cyg} given by Najarro et al. (1997), which has the main 
parameters $T_{eff}$ = 18\,200 K, ${\log}{\,}{L/L_{\odot}}$ = 5.75
and $n_{He}/n_{H} =$ 0.3. The physical conditions at the level where 
the short-period absorption components are formed were derived by 
using the observation 
({\em cf.} Fig. 1) that these absorption components are formed at the 
level where 
the stellar wind velocity is between 30 and 100 
${\rm km}{\,}{{\rm s}^{-1}}$. We took an average value of 60 
${\rm km}{\,}{{\rm s}^{-1}}$. From Najarro et al.'s velocity law
we then find that this level corresponds with $r = 2.1{\,}R_*$. The 
density and pressure at that level follow from Najarro et al.'s 
rate of mass loss. We assumed for that level a temperature of 12,000 K,
an assumption that apeared not to be critical. That yields the 
continuous absorption coefficient. The value of ${\Gamma}_1$ was 
computed using a non-LTE approach involving the 16 most abundant 
elements, developed by Lobel ({\em in prep}). 

At that level $R_* = 5.24 {\: 10^{12}}$ cm, $g_{\rm rad} = -22.49 
{\: \rm cm}{\:}{\rm s}^{-2}$ ({\em cf.} Lamers and Cassinelli, 1999) and
$\delta (\delta w / \delta r)  = -3.626 {\: 10^{-20}}{\: \rm s}^{-1}$. 
Hence $F_2 = -2.152 10^{8}{\: \rm cm}{\:}{\rm s}^{-1}$.   

The two asymptotic cases are easiest to handle. With $k_z = 0$, 
hence assuming 
strictly horizontal waves, and taking  
$P = 17.3$ d we obtain for the horizontal wavelength 
$L_x = 1.982 {\: 10^{12}}$ cm or $0.38 R_*$. For purely
vertical waves ($k_x = 0$) we get $L_z = 1.22 {\: 10^{10}}$ cm or 
$0.0023{\:} R_*$.   

The waves are gravity waves since for $P = 17.3$ d the corresponding
value 
of $\omega$ falls below the Brunt-V{\"{a}}is{\"{a}}l{\"{a}} frequency 
(de Jager, 2001). Since 
gravity waves mostly travel horizontally we tend to prefer the first 
solution, that
for horizontal waves. An additional argument for choosing that option 
is that 
the wavelengh ( 0.38$R_*$) is close to the wavelength suggested by the
observations, for which we concluded in Section 2 that the waves should 
have 
lengths of the order of the stellar radius.   
  
The {\em conclusion} of this Section is that if the short-period 
absorption
components would have a period of 17.3 d, then they could be interpreted
as 
horizontally propagating gravity waves with wavelengths of the order 
of 0.4
times the stellar radius.   

\section {Conclusions}

The high S/N and high-resolution observations reported here show 
for the first time the existence of short-period narrow absorption 
components in the cores of 
high Balmer lines and their rapid variations. If their period was
17.3 d, which is the only known short period observed in the light of 
\astrobj{P Cyg}, they should be interpreted as gravity waves with wavelengths 
of the order of 0.4 times the stellar radius, a conclusion that is also 
suggested by the observed behaviour of the absorption components, 
which seem to consist of 
two waves in counterphase. A period differing by a factor 2 or so 
would not 
change these conclusions. But, evidently, more observations are 
needed to check this statement and we urge observers to monitor these 
features. 

\vspace{0.3cm}

We thank M. Friedjung, A. van Genderen, A. de Koter, O. Stahl and R. 
Stothers for
help and useful remarks. Many thanks are due to H. Nieuwenhuijzen for 
advice and efficient help with part of the computations and Ilia Ilyin
for help with SOFIN data reduction.

{}

\end{document}